\documentclass{desyproc}

\def \et {E_{T}}
\newcommand{\met}{\mbox{$\not\!\!\et$}}

\newcommand{\be}{\begin{equation}}
\newcommand{\ee}{\end{equation}}
\newcommand{\bea}{\begin{eqnarray}}
\newcommand{\eea}{\end{eqnarray}}

\begin{document}
\title{Physics with tagged protons at the LHC: understanding the Pomeron
structure and anomalous coupling studies}

\author{{\slshape Christophe Royon}\\[1ex]
IRFU-SPP, CEA Saclay, F91 191 Gif-sur-Yvette cedex, France
}

\contribID{smith\_joe}


\acronym{EDS'13} 

\maketitle

\begin{abstract}
We describe different physics topics which can be performed at the LHC using
tagged intact protons, namely a better understanding of the Pomeron structure in
terms of quarks and gluons, and the serach for quartic anomalous couplings.
\end{abstract}

 \section{Inclusive diffraction measurement at the LHC}
In this section, we discuss potential measurements at the LHC that can constrain
the Pomeron structure. The Pomeron structure in terms of quarks and gluons has been
derived from QCD fits at HERA and at the Tevatron and it is possible to probe this structure and the
QCD evolution at the LHC in a completely new kinematical domain.

\subsection{Dijet production in double Pomeron exchanges processes}

The high energy and luminosity at the LHC allow the exploration of a 
completely new kinematical domain. One can first probe if the Pomeron is universal between
$ep$ and $pp$ colliders, or in other
other words, if we are sensitive to the same object at HERA and the LHC. 
The different diagrams of the processes that can be studied at the LHC
are namely double pomeron exchange (DPE) production of dijets,
of $\gamma +$jet, sensitive respectively to the gluon and quark contents of the
Pomeron, and the jet gap jet events. All diagrams were included in the FPMC~\cite{FPMC}
generator that was used for this analysis.

\begin{figure}
\epsfig{file=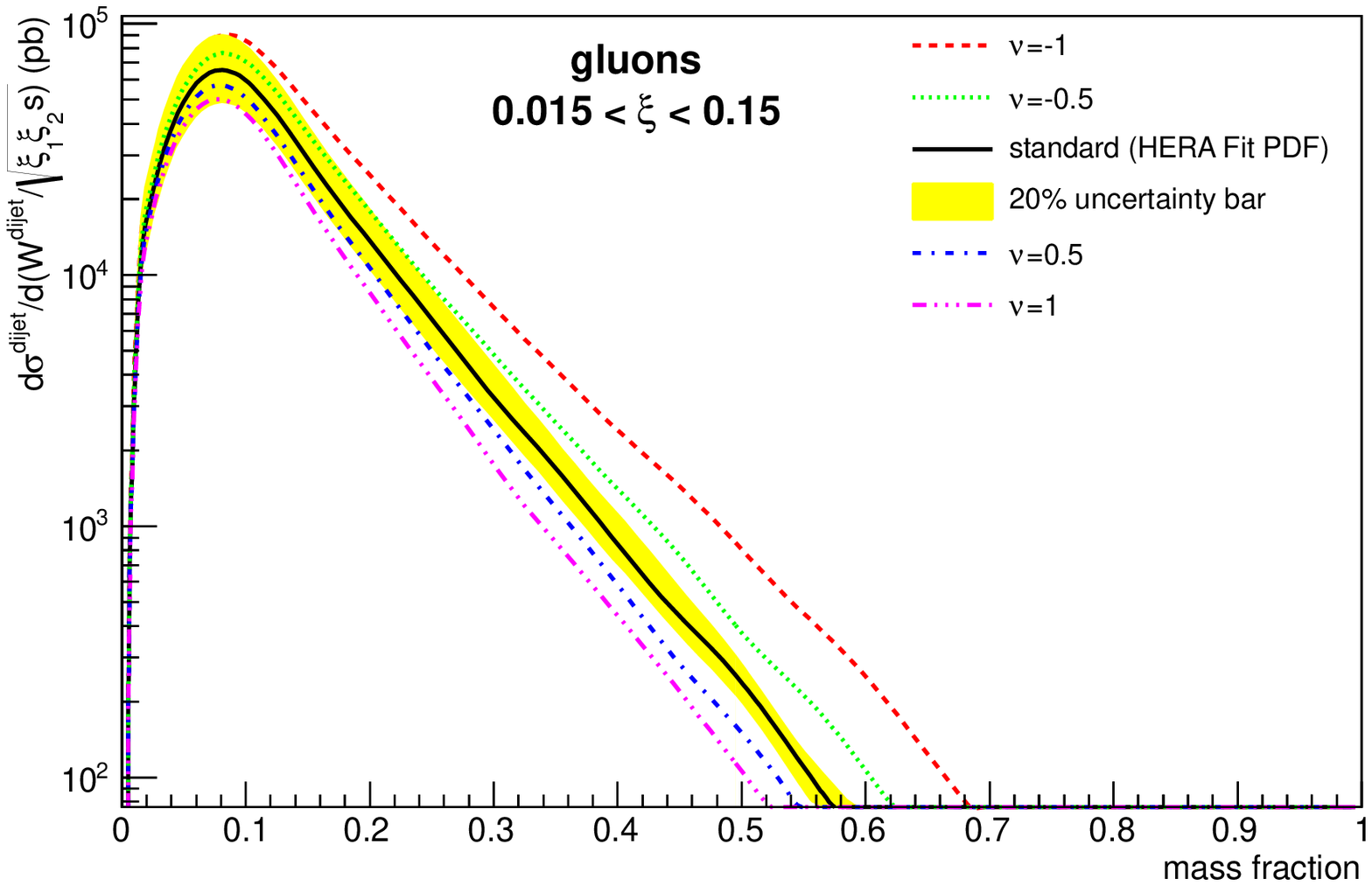,width=6.5cm}
\epsfig{file=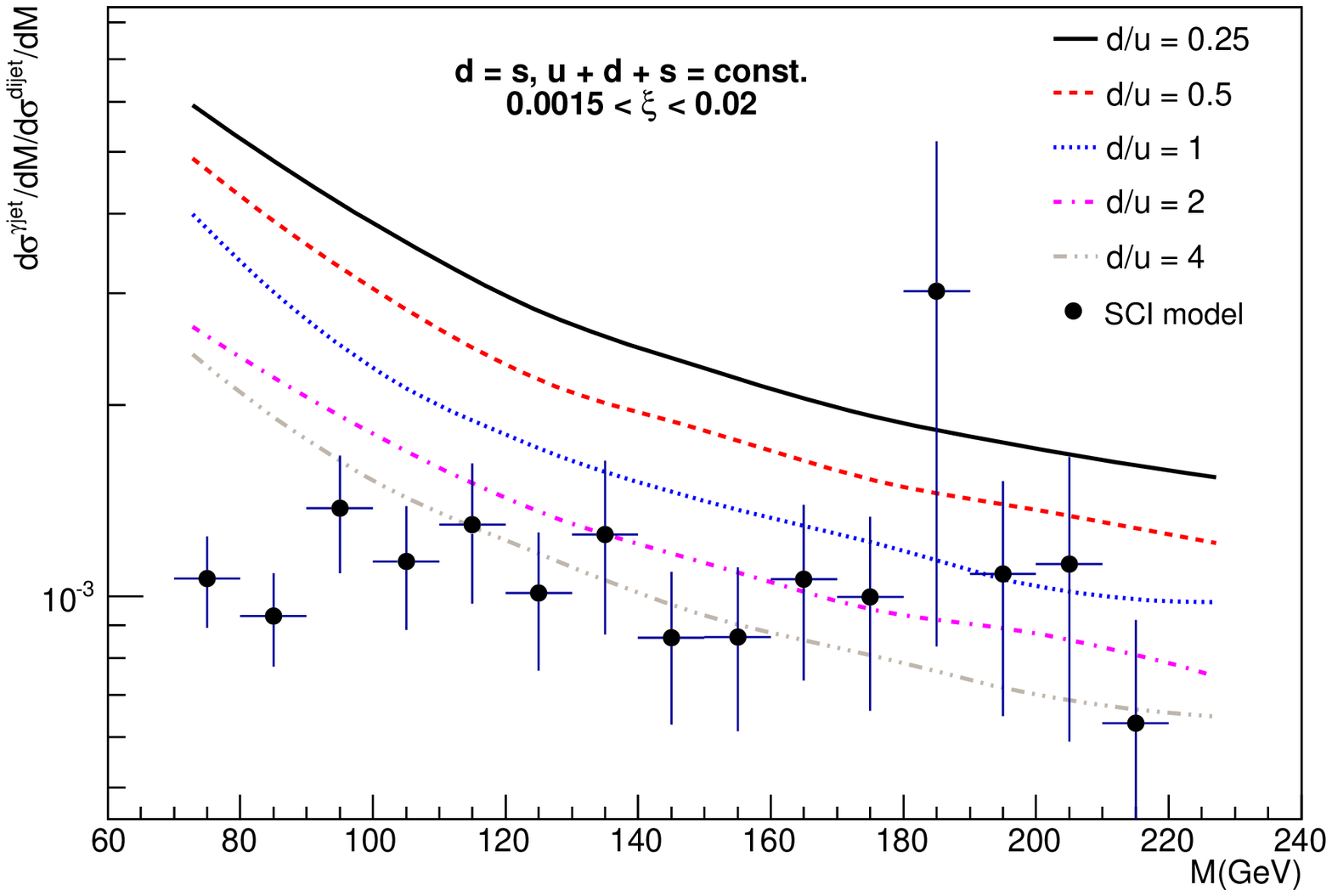,width=6.5cm}

\caption{Left: DPE di-jet mass fraction distribution. The different curves correspond to different 
modifications of the Pomeron gluon density extracted from HERA data (see text).
Right: DPE $\gamma +$ jet to di-jet differential cross section ratio, for the 
acceptance of the 210m proton detectors as a function of the diffractive mass.}
\label{d0}
\end{figure}

The dijet production in DPE events at the LHC is sensitive to the gluon density
in the Pomeron. In order to
quantify how well we are sensitive to the Pomeron structure in terms of gluon
density at the LHC, we display in Fig.~\ref{d0}, left, the dijet cross
section as a function of the dijet mass fraction~\cite{matthias}, assuming for instance
the protons to be tagged in the AFP~\cite{afp} proton detectors at 210 m. The central black line displays the cross
section value for the gluon density in the Pomeron measured at HERA including an
additional survival probability of 0.03. The yellow band shows the effect of the
20\% uncertainty on the gluon density taking into account the normalisation
uncertainties. The dashed curves display how the dijet
cross section at the LHC is sensitive to the gluon density distribution
especially at high $\beta$. For this sake, we multiply the gluon density in the
Pomeron from HERA by $(1-\beta)^{\nu}$ where $\nu$ varies between -1 and 1. When
$\nu$ is equal to -1 (resp. 1), the gluon density is enhanced (resp, decreased)
at high $\beta$. From Fig.~\ref{d0}, we notice that the dijet cross section
is indeed sensitive to the gluon density in the Pomeron 
and we can definitely check if the Pomeron
model from HERA and its structure in terms of gluons is compatible between HERA
and the LHC. This will be an important test of the Pomeron universality. This
measurement can be performed for a luminosity as low as 10 pb$^{-1}$ since the
cross section is very large (typically, one day at low luminosity without pile
up at the LHC). 

\subsection{Sensitivity to the Pomeron structure in quarks using $\gamma + \textnormal{jet}$ events}

Fig.~\ref{d0}, right, displays a possible observable at the LHC that can 
probe the
quark content in the Pomeron, namely the $\gamma +$jet to the
dijet cross section ratio~\cite{matthias} as a function of diffractive mass ($\sqrt{\xi_1 \xi2
S}$) for different
assumptions on the quark content of the Pomeron, $d/u$ varying between 0.25 and
4 in steps of 0.25. We notice that the cross section ratio varies by a factor
2.5 for different values of $u/d$. 
The aim of the diffractive mass distribution measurement is twofolds: is the Pomeron 
universal between
HERA and the LHC and what is the quark content of the Pomeron? The QCD 
diffractive fits at HERA assumed that
$u=d=s=\bar{u}=\bar{d}=\bar{s}$, since data were not sensitive to the
difference between the different quark component in the Pomeron. 
The LHC data will allow us to determine for instance which value of $d/u$ is
favoured by data. Let us assume that $d/u=0.25$ is favoured. If this is the
case, it will be needed to go back to the HERA QCD diffractive fits and check if
the fit results at HERA can be modified to take into account this assumption. If
the fits to HERA data lead to a large $\chi^2$, it would indicate that the Pomeron is not
the same object at HERA and the LHC. On the other hand, if the HERA fits work
under this new assumption, the quark content in the Pomeron will be further
constrained. The advantage of measuring the cross section ratio as a function of
diffractive mass is that most of the systematic uncertainties will cancel.

Soft color interaction models~(SCI) is another model to explain diffraction at
hadronic colliders~\cite{Edin:1995gi}. In Fig.~\ref{d0}, right, we notice that
the distribution of the $\gamma+$jet to dijet ratio as a function of the total
diffractive mass distributions
may allow to distinguish between the Herwig/DPE and Pythia/SCI models
because the latter leads to a more flat dependence on the total diffractive mass,
giving further insight into soft QCD.

\subsection{Jet gap jet production in double Pomeron exchanges processes}
In this process, both protons are intact 
after the interaction and detected in AFP at 210 m, two jets are measured in the 
ATLAS central detector and a gap devoid of any energy is present between 
the two jets~\cite{jgjpap}. This kind of event is important since it is sensitive to QCD 
resummation dynamics given by the BFKL~\cite{bfkl,fwd} 
evolution equation. This process has never been measured to date and will be 
one of the best methods to probe these resummation effects, benefitting from 
the fact that one can perform the measurement for jets separated by a large 
angle (there is no remnants which `pollute' the event). As an example, the 
cross section ratio for events with gaps to events with or without gaps 
as a function of the leading jet $p_T$  is of the order of 20\% which is
much higher than the expectations for non-diffractive events. This is due to the fact that the
survival probability of 0.03 at the LHC does need to be applied for diffractive events.

\section{Exclusive $WW$ and $ZZ$ production}

In the Standard Model (SM) of particle physics, the couplings of fermions and 
gauge bosons are constrained by the gauge symmetries of the Lagrangian.
The measurement of $W$ and $Z$ boson pair productions via the exchange of
two photons  
allows to provide directly stringent tests
of one of the most important and least understood
mechanism in particle physics, namely the
electroweak symmetry breaking~\cite{us}.

The parameterization of the quartic couplings
based on \cite{Belanger:1992qh} is adopted. 
The cuts to select quartic anomalous gauge coupling $WW$ events are the following, 
namely $0.0015<\xi<0.15$ for the
tagged protons corresponding to the AFP detector at 210 and 420 m, $\met>$ 20 GeV, 
$\Delta \phi<3.13$ between the two leptons. In
addition, a cut on the $p_T$ of the leading lepton $p_T>160$ GeV and on the
diffractive mass $W>800$ GeV are requested since anomalous coupling events
appear at high mass.
After these requirements, we expect about 0.7 background
events for an expected signal of 17 events if the anomalous coupling is about
four orders of magnitude lower than the present LEP limit~\cite{opal} ($|a_0^W / \Lambda^2| =
5.4$ 10$^{-6}$) for a luminosity of 30 fb$^{-1}$, and about two orders of
magnitude better than the present CMS limits~\cite{cms}.

The search for quartic anomalous couplings between $\gamma$ and $W$ bosons was
performed again after a full simulation of the ATLAS detector including pile
up~\cite{afp} assuming the protons to be tagged in AFP at 210 m only. 
Integrated luminosities of 40 and 300 fb$^{-1}$ with, 
respectively, 23 or 46 average pile-up
events per beam crossing have been considered. In order to reduce the
background, each $W$  
is assumed to decay leptonically (note that the semi-leptonic case in under study). 
The full list of background processes 
used for the ATLAS measurement of Standard Model $WW$ cross-section was
simulated, namely $t \bar{t}$, $WW$, $WZ$, $ZZ$, $W+$jets, Drell-Yan and 
single top events. In addition, the additional diffractive backgrounds
were also simulated,
Since only leptonic decays of the 
W bosons are considered, we require in addition less than 3 tracks associated 
to the primary vertex, which allows us to reject a large fraction of the
non-diffractive backgrounds (e.g. $t \bar{t}$, diboson
productions, $W+$jet, etc.) since they show much higher track multiplicities. 
Remaining Drell-Yan and
QED backgrounds are suppressed by requiring the difference in azimuthal angle between the
two leptons $\Delta \phi <$ 3.1.  After these requirements, a similar
sensitivity with respect to fast simulation without pile up was obtained. 

Of special interest will be also the search for anomalous quartic $\gamma \gamma \gamma
\gamma$ anomalous couplings which is now being implemented in the FPMC
generator. Let us notice that there is no present existing limit on such
coupling and the sensitivity using the forward proton detectors is expected to
be similar as the one for $\gamma \gamma WW$ or $\gamma \gamma ZZ$ anomalous
couplings. If discovered at the LHC, $\gamma \gamma \gamma
\gamma$ quartic anomalous couplings might be related to the existence of
extra-dimensions in the universe, which might lead to a reinterpretation of
some experiments in atomic physics. As an example, the Aspect photon correlation
experiments~\cite{aspect} might be interpreted via the existence of 
extra-dimensions. Photons
could communicate through extra-dimensions and the deterministic interpretation
of Einstein for these experiments might be true if such anomalous
couplings exist. From the point of view of atomic physics, the results of the
Aspect experiments would depend on the distance of the two photon sources.


\begin{footnotesize}

\end{footnotesize}

\begin{thebibliography}{99}

\bibitem{FPMC}  
M. Boonekamp, A. Dechambre, V. Juranek, O. Kepka, M. Rangel, 
C. Royon, R. Staszewski, e-Print: arXiv:1102.2531;    
M. Boonekamp, V. Juranek, O. Kepka, C. Royon ``Forward Physics Monte Carlo'',
      ``Proceedings of the workshop: HERA and the LHC workshop series on the
      implications of HERA for LHC physics,'' arXiv:0903.3861 [hep-ph].


\bibitem{matthias} C, Marquet, C. Royon, M. Saimpert, D. Werder, 
arXiv:1306.4901, Phys.Rev. D{\bf 87} (2013) 3, 034010.



\bibitem{afp} ATLAS Coll., CERN-LHCC-2011-012.



\bibitem{Edin:1995gi}
  A.~Edin, G.~Ingelman and J.~Rathsman,
  Phys.\ Lett.\ B {\bf 366}, 371 (1996); Z.Phys. C{\bf 75}, 57 (1997);
  J.~Rathsman,
  Phys.\ Lett.\  B {\bf 452} (1999) 364.

.
\bibitem{jgjpap}	
C. Marquet, C. Royon, M. Trzebinski, R. Zlebcik, 
Phys.Rev. D{\bf 87} (2013) 3, 034010;
O. Kepka, C. Marquet, C. Royon,
Phys. Rev. D{\bf 83} (2011) 034036.


\bibitem{bfkl} V. S. Fadin, E. A. Kuraev, L. N. Lipatov, Phys. Lett. B{\bf 60} (1975) 50; I. I. Balitsky, L. N. 
Lipatov, Sov.J.Nucl.Phys. {\bf 28} (1978) 822;

.

\bibitem{fwd} C. Marquet, C. Royon, Phys. Rev. D{\bf 79} (2009) 034028; 
O. Kepka, C. Royon, C. Marquet, R. Peschanski, Eur. Phys .J. C{\bf 55} (2008) 
259-272; Phys. Lett. B{\bf 655} (2007) 236-240; 
H. Navelet, R. Peschanski, C. Royon, S. Wallon, Phys. Lett. B{\bf 385} (1996)
357;H. Navelet, R. Peschanski, C. Royon, Phys. Lett. B366 (1996) 329. 


\bibitem{us} E. Chapon, O. Kepka, C. Royon, Phys. Rev. D{\bf 81} (2010) 074003;
O.~Kepka and C.~Royon,
Phys.\ Rev.\  D {\bf 78} (2008) 073005;
J. de. Favereau et al., preprint arXiv:0908.2020.

\bibitem{Belanger:1992qh}
  G.~Belanger and F.~Boudjema,
  Phys.\ Lett.\  B {\bf 288} (1992) 201.

\bibitem{opal}
  G.~Abbiendi {\it et al.}  [OPAL Collaboration],
  Phys.\ Rev.\  D {\bf 70} (2004) 032005
  [arXiv:hep-ex/0402021].
  
\bibitem{cms} CMS Coll., JHEP {\bf 07} (2013) 116.
 
\bibitem{aspect} A. Aspect, P. Grangier, G. Roger, Phys. Rev. Lett., Vol. 49, no
2 (1982) 91-94;
A. Aspect, J. Dalibard, G. Roger, Phys. Rev. Lett., Vol. 49, Iss. 25 (1982)
1804.






\end{thebibliography}
\end{document}